\documentstyle[twoside,fleqn,espcrc2,epsf,rotate]{article}

\newcommand{\AmS}{{\protect\the\textfont2
  A\kern-.1667em\lower.5ex\hbox{M}\kern-.125emS}}

\def\myfigure#1{\centerline{#1}}
\def\Preprint{\vspace*{-7.5cm} %\noindent hep-ph/9612XXX
 \noindent FTUV/00-1214 \\ %\mbox{}\hfill
  IFIC/00-85 \\  %\mbox{}\hfill December 1996 \\
  \vspace{5.3cm}}
\def\refjl#1#2#3#4#5#6{\bibitem{#1} #2, {#3} {#4} (#5) #6.}

\def\etal{{et al}}
\def\NP{Nucl. Phys.}
\def\NPPS{Nucl. Phys. B (Proc. Suppl.)}
\def\PL{Phys. Lett.}
\def\PRL{Phys. Rev. Lett.}
\def\PR{Phys. Rev.}

\def\ZP{Z. Phys.}

\def\JPG{J. Phys. G: Nucl. Phys.}           %1975--1988
\def\RPP{Rep. Prog. Phys.}

\def\PPNP{Prog. Part. Nucl. Phys.}

\newcommand{\eqn}[1]{(\ref{#1})}
\newcommand{\be}{\begin{equation}}
\newcommand{\ee}{\end{equation}}
\newcommand{\no}{\nonumber}
\newcommand{\bel}[1]{\be\label{#1}}
\newcommand{\ba}{\begin{array}{c}}
\newcommand{\bat}{\begin{array}{cc}}
\newcommand{\ea}{\end{array}}
\newcommand{\beqn}{\begin{eqnarray}}
\newcommand{\eeqn}{\end{eqnarray}}

\newcommand{\bi}{\begin{itemize}}
\newcommand{\ei}{\end{itemize}}

\newcommand{\rms}{\rm\scriptsize}

\newcommand{\lsim}{~{}_{\textstyle\sim}^{\textstyle <}~}

\newcommand{\cO}{{\cal O}}
\newcommand{\cP}{{\cal P}}

\newcommand{\cA}{{\cal A}}

\title{Tau Physics: Theoretical Perspective  %\thanks{
       }

\author{A. Pich\address{      %\\ \noindent  %\address{
         Departament de F\'{\i}sica Te\`orica,
         IFIC, Universitat de Val\`encia --- CSIC, \\
         Apt. Correus 22085, E--46071 Val\`encia, Spain}}
    
\begin{document}

\begin{abstract}
The leptonic decays of the $\tau$ lepton provide relevant
tests on the structure of the
weak currents  and the universality of their couplings to the gauge
bosons. The hadronic $\tau$ decay modes constitute an
ideal  tool for  studying low--energy effects of the strong
interaction in  very clean conditions.
Accurate determinations of the QCD coupling and the strange quark mass
have been obtained with $\tau$ decay data.
New physics phenomena, such as a non-zero $m_{\nu_\tau}$ or
violations of conservation laws
can also be searched for with $\tau$ decays.
\end{abstract}


\maketitle
\Preprint
\section{INTRODUCTION}
\label{sec:introduction}

The $\tau$ lepton is a member of the third generation which decays
into particles belonging to the first and second ones.
Thus, $\tau$ physics could provide some
clues to the puzzle of the recurring families of leptons and quarks.
One na\"{\i}vely expects the heavier fermions to be more sensitive
to whatever dynamics is responsible for the fermion--mass generation.
The leptonic or semileptonic character of $\tau$  decays
provides a clean laboratory to test the structure of the weak
currents  and the universality of their couplings to the gauge bosons.
Moreover, the  $\tau$ is
the only known lepton massive enough to  decay  into  hadrons;
its  semileptonic decays are an ideal tool for studying
strong interaction effects in  very clean conditions.

The last few years have witnessed \cite{tau98,taurev,Stahl}
a substantial change on our knowledge of the $\tau$ properties.
The large (and clean) data samples collected by the most recent experiments
have improved considerably the statistical accuracy and, moreover,
have brought a new level of systematic understanding.
On the theoretical side, the detailed study of higher--order
electroweak corrections and QCD contributions has promoted the physics
of the $\tau$ lepton to the level of precision tests.

\section{UNIVERSALITY}
\label{sec:universality}

\subsection{Charged Currents}
\label{subsec:cc}

%%%%%%%%%%%%%%% Table Tau Properties %%%%%%%%%%%%%%%%%%%%%%%%%%%%%%
%
\begin{table}[thb]
\centering
\caption{Average values \protect\cite{RO:00,PDG:00,OPAL:00}
of some basic $\tau$ parameters.}
\label{tab:parameters}
\vspace{0.2cm}
\begin{tabular}{lc}
\hline
$m_\tau$ & $(1777.03^{+0.30}_{-0.26})$ MeV \\
$\tau_\tau$ & $(290.89\pm 1.00)$ fs \\
$B_e$ & $(17.804\pm 0.051)\% $ \\
$B_\mu$ & $(17.336\pm 0.051)\% $ \\
%$R_\tau^B \equiv (1-B_e-B_\mu)/ B_e$ & $3.645\pm 0.022$ \\
Br($\tau^-\to\nu_\tau\pi^-$) & $(11.03\pm 0.14)\% $ \\
Br($\tau^-\to\nu_\tau K^-$) & $(0.684\pm 0.022)\% $ \\
\hline
\end{tabular}
\end{table}
%
%%%%%%%%%%%%%%%%%%%%%%%%%%%%%%%%%%%%%%%%%%%%%%%%%%%%%%%%%%%%%%%%%%%%%

The leptonic decays
$\tau^-\to e^-\bar\nu_e\nu_\tau,\mu^-\bar\nu_\mu\nu_\tau$
are theoretically understood at the level of the electroweak
radiative corrections \cite{MS:88}.
Within the Standard Model (SM),
\begin{equation}
\label{eq:leptonic}
%\Gamma_{\tau\to l} \, = \,
\Gamma (\tau^- \rightarrow \nu_{\tau} l^- \bar{\nu}_l)  \, = \,
  {G_F^2 m_{\tau}^5 \over 192 \pi^3} \, f(m_l^2 / m_{\tau}^2) \,
r_{EW},
\end{equation}
where
%$\Gamma_{\tau\to l}\equiv
%\Gamma (\tau^- \rightarrow \nu_{\tau} l^- \bar{\nu}_l)$,
$f(x) = 1 - 8 x + 8 x^3 - x^4 - 12 x^2 \log{x}$.
The factor $r_{EW}=0.9960$ takes into account radiative corrections
not included in the
Fermi coupling constant $G_F$, and the non-local structure of the
$W$ propagator \cite{MS:88}.

%%%%%%%%%%%%% FIGURE 1 %%%%%%%%%%%%%%%%%
\begin{figure}[bth]
\myfigure{\epsfxsize =7.5cm \epsfbox{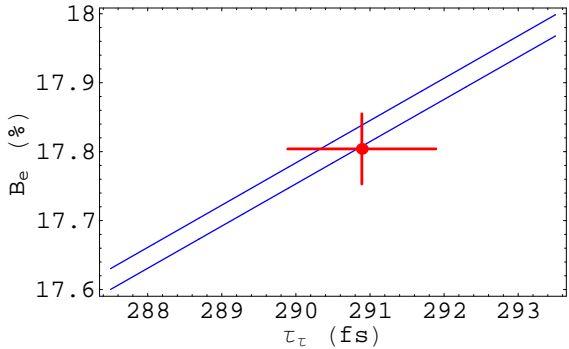}}
\vspace{-0.5cm}
\caption{Relation between $B_e$ and $\tau_\tau$. The
band corresponds to the prediction in Eq.~(\protect\ref{eq:relation}).
\label{fig:BeLife}}
\end{figure}
%%%%%%%%%%%%%%%%%%%%%%%%%%%%%%%%%%%%%%%%%

Using the value of $G_F$ measured in $\mu$ decay,
$G_F = (1.16637 \pm 0.00001)\times 10^{-5}\:\mbox{\rm GeV}^{-2}$
\cite{MS:88,RS:99}, Eq.~\eqn{eq:leptonic} provides a
relation \cite{taurev} between the leptonic branching ratios
$B_l\equiv B(\tau^-\to\nu_\tau l^-\bar\nu_l)$ and the
$\tau$ lifetime:
\beqn
\label{eq:relation}
B_e & = & {B_\mu \over 0.972564\pm 0.000010}
\no\\ & = &
{ \tau_{\tau} \over (1.6321 \pm 0.0014) \times 10^{-12}\, {\rm s} } \, .
\eeqn
The errors reflect the present uncertainty of $0.3$ MeV
in the value of $m_\tau$.

%%%%%%%%%%%%%%%%  TABLE  %%%%%%%
\begin{table}[bth]
\centering
\caption{Present constraints on $|g_l/g_{l'}|$.}
\label{tab:ccuniv}
\vspace{0.2cm}
\begin{tabular}{lc}
\hline
& $|g_\mu/g_e|$ \\ \hline
$B_\mu/B_e$ & $1.0006\pm 0.0021$ \\
$B_{\pi\to e}/B_{\pi\to\mu}$ & $1.0017\pm 0.0015$ \\
% $\sigma\cdot B_{W\to\mu/e}$\  ($p\bar p$) & $0.98\pm 0.03$\\
$B_{W\to\mu}/B_{W\to e}$  & $0.999\pm 0.013$ \\
\hline\hline
& $|g_\tau/g_\mu|$  \\ \hline
$B_e\tau_\mu/\tau_\tau$ & $0.9995\pm 0.0023$ \\
$\Gamma_{\tau\to\pi}/\Gamma_{\pi\to\mu}$ &  $1.005\pm 0.007$ \\
$\Gamma_{\tau\to K}/\Gamma_{K\to\mu}$ & $0.977\pm 0.016$ \\
$B_{W\to\tau}/B_{W\to\mu}$  & $1.022\pm 0.014$
\\ \hline\hline
& $|g_\tau/g_e|$  \\ \hline
$B_\mu\tau_\mu/\tau_\tau$ & $1.0001\pm 0.0023$ \\
%  $\sigma\cdot B_{W\to\tau/e}$\  ($p\bar p$) & $0.987\pm 0.025$\\
$B_{W\to\tau}/B_{W\to e}$  & $1.021\pm 0.015$
\\ \hline
\end{tabular}
\end{table}
%
%%%%%%%%%%%%%%%%%%%%%%%%%%%%%%%%%%%%%%%%%%%%%%%%

The relevant experimental measurements are given in Table~\ref{tab:parameters}.
The predicted $B_\mu/B_e$ ratio is in perfect agreement with the measured
value $B_\mu/B_e = 0.974 \pm 0.004$.  As shown in
Fig.~\ref{fig:BeLife}, the relation between $B_e$ and
$\tau_\tau$ is also well satisfied by the present data.
The experimental precision (0.3\%) is already approaching the
level where a possible non-zero $\nu_\tau$ mass could become relevant; the
present bound \cite{ALEPHnumass}
$m_{\nu_\tau}< 18.2$ MeV (95\% CL) only guarantees that such effect
is below 0.08\%.

These measurements can be used to test the universality of
the $W$ couplings to the leptonic charged currents.
The $B_\mu/B_e$ ratio constraints $|g_\mu/g_e|$, while
$B_e/\tau_\tau$ and $B_\mu/\tau_\tau$
provide information on $|g_\tau/g_\mu|$ and $|g_\tau/g_e|$, respectively.
The present results are shown in Table~\ref{tab:ccuniv},
together with the values obtained from the ratios
$\Gamma(\pi^-\to e^-\bar\nu_e)/\Gamma(\pi^-\to\mu^-\bar\nu_\mu)$
\cite{BR:92} and
$\Gamma(\tau^-\to\nu_\tau P^-)/
\Gamma(P^-\to \mu^-\bar\nu_\mu)$ \  [$P=\pi,K$].
Also shown are the direct constraints obtained from the
$W^-\to l^-\bar\nu_l$ decay modes at LEP II \cite{LEP:99}.
The present data verify the universality of the leptonic
charged--current couplings to the 0.15\% ($e/\mu$) and 0.23\%
($\tau/\mu$, $\tau/e$) level.

\subsection{Neutral Currents}
\label{subsec:nc}

%%%%%%%%%%%%%%%%   FIGURE LEP %%%%%%%%%%%%%%%%
\begin{figure}[tbh]
\myfigure{\epsfxsize =7.5cm \epsfbox{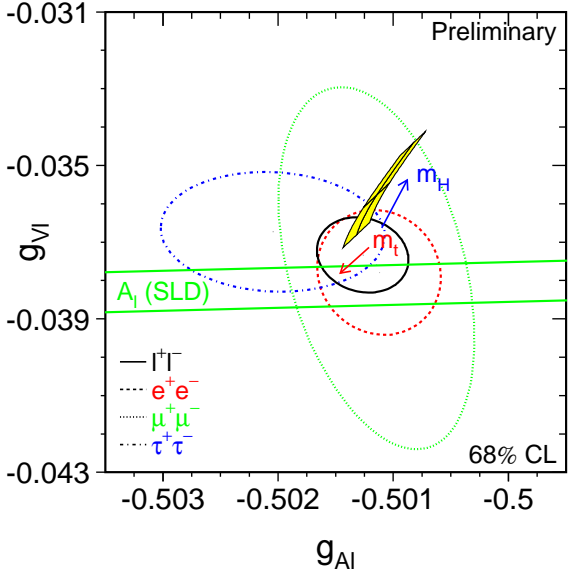}}
\vspace{-0.5cm}
\caption{68\% probability contours in the $a_l$-$v_l$ plane
from LEP data \protect\cite{LEP:99}.
%The solid contour assumes lepton universality.
Also shown is the $1\sigma$ band resulting from the
$\protect\cA_{\mbox{\protect\rms LR}}^0$ measurement at SLD.
The grid corresponds to the SM prediction.}
\label{fig:gagv}
\end{figure}
%%%%%%%%%%%%%%%%%%%%% END FIGURE %%%%%%%%%%%%%%%%%%%%%%%%%%

In the SM, all leptons with equal electric charge have identical
couplings to the $Z$ boson.
This has been tested at LEP and SLC \cite{LEP:99},
by measuring the total $e^+e^-\to Z \to l^+l^-$
cross--section, the forward--backward asymmetry,
the (final) polarization asymmetry, the forward--backward (final)
polarization
asymmetry, and (at SLC) the left--right asymmetry between the
cross--sections for initial left-- and right--handed electrons
and the left--right forward--backward asymmetry.
The $Z$ partial decay width to the $l^+l^-$ final state
determines the sum $(v_l^2 + a_l^2)$, while the ratio $v_l/a_l$
is derived from the asymmetries, which measure the average longitudinal
polarization of the lepton $l^-$:
$\cP_l \equiv  - 2 v_l a_l / (v_l^2 + a_l^2)$.

The measurement of the final polarization asymmetries can (only) be
done for $l=\tau$, because the spin polarization of the $\tau$'s
is reflected in the distorted distribution of their decay products.
Therefore, $\cP_\tau$ and $\cP_e$ can be determined from a
measurement of the spectrum of the final charged particles in the
decay of one $\tau$, or by studying the correlated distributions
between the final products of both $\tau's$ \cite{ABGPR:92,DDDR:93}. %BPR:91

The data are in excellent agreement with the SM predictions
and confirm the universality of the leptonic neutral couplings.
Figure~\ref{fig:gagv} shows the 68\% probability contours in
the $a_l$--$v_l$ plane, obtained from a
combined analysis \cite{LEP:99} of all leptonic observables.

\section{LORENTZ STRUCTURE}
\label{sec:lorentz}

Let us consider the leptonic decay $l^-\to\nu_l l'^-\bar\nu_{l'}$.
The most general, local, derivative--free, lepton--number conserving,
four--lepton interaction Hamiltonian,
consistent with locality and Lorentz invariance
\cite{MI:50,FGJ:86},
\be
{\cal H} = 4 \frac{G_{l'l}}{\sqrt{2}}
\sum_{n,\epsilon,\omega}          %^{n = S,V,T}
g^n_{\epsilon\omega}   %g^n_{l'_\epsilon l^{\phantom{'}}_\omega}
\left[ \overline{l'_\epsilon}
\Gamma^n {(\nu_{l'})}_\sigma \right]\,
\left[ \overline{({\nu_l})_\lambda} \Gamma_n
	l_\omega \right]\ ,
\label{eq:hamiltonian}
\ee
contains ten complex coupling constants or, since a common phase is
arbitrary, nineteen independent real parameters
which could be different for each leptonic decay.
The subindices
$\epsilon , \omega , \sigma, \lambda$ label the chiralities (left--handed,
right--handed)  of the  corresponding  fermions, and $n$ the
type of interaction:
scalar ($I$), vector ($\gamma^\mu$), tensor
($\sigma^{\mu\nu}/\sqrt{2}$).
For given $n, \epsilon ,
\omega $, the neutrino chiralities $\sigma $ and $\lambda$
are uniquely determined.

The total decay width is proportional to the following combination
of couplings, which is usually normalized to one \cite{FGJ:86}:
\beqn\label{eq:normalization}
1 &\!\!\! = &\!\!\!
{1\over 4} \,\left( |g^S_{RR}|^2 + |g^S_{RL}|^2
    + |g^S_{LR}|^2 + |g^S_{LL}|^2 \right)
\no \\ & &\!\!\! \mbox{}
+ \left(
   |g^V_{RR}|^2 + |g^V_{RL}|^2 + |g^V_{LR}|^2 + |g^V_{LL}|^2 \right)
\no\\ & &\!\!\! \mbox{}
   +  3 \,\left( |g^T_{RL}|^2 + |g^T_{LR}|^2 \right)
\no\\ & \equiv &\!\!\! Q_{RR} + Q_{RL} + Q_{LR} +Q_{LL}
\, .
\eeqn
The universality tests mentioned before refer then to the global
normalization $G_{l'l}$, while the $g^n_{\epsilon \omega}$ couplings
parameterize the relative strength of different types of interaction.
The sums $Q_{\epsilon \omega}$ of all factors with the
same subindices give the probability of having a decay from an
initial charged lepton with chirality $\omega$ to a final one with
chirality $\epsilon$.
In the SM, $g^V_{LL}  = 1$  and all other
$g^n_{\epsilon\omega} = 0 $.

The energy spectrum and angular distribution of the
final charged lepton provides information on
the couplings $g^n_{\epsilon \omega}$.
For $\mu$ decay, where precise measurements of the polarizations of
both $\mu$ and $e$ have been performed, there exist \cite{FGJ:86}
upper bounds on $Q_{RR}$, $Q_{LR}$ and $Q_{RL}$, and a lower bound
on $Q_{LL}$. They imply corresponding upper bounds on the 8
couplings $|g^n_{RR}|$, $|g^n_{LR}|$ and $|g^n_{RL}|$.
The measurements of the $\mu^-$ and the $e^-$ do not allow to
determine $|g^S_{LL}|$ and $|g^V_{LL}|$ separately \cite{FGJ:86,JA:66}.
Nevertheless, since the helicity of the $\nu_\mu$ in pion decay is
experimentally known \cite{FE:84}
to be $-1$, a lower limit on $|g^V_{LL}|$ is
obtained \cite{FGJ:86} from the inverse muon decay
$\nu_\mu e^-\to\mu^-\nu_e$. These limits show nicely
that the bulk of the $\mu$--decay transition amplitude is indeed of
the predicted V$-$A type:
$|g^V_{LL}| > 0.960$ \ (90\% CL) \cite{PDG:00}.
Improved bounds on the $\mu$ couplings are expected from
the Twist experiment \cite{Rodning} at TRIUMF.

%%%%%%%%%%%%  FIGURE  Tau Couplings %%%%%%%%%%
\begin{figure}[tbh]
\myfigure{\epsfxsize =6.75cm \epsfbox{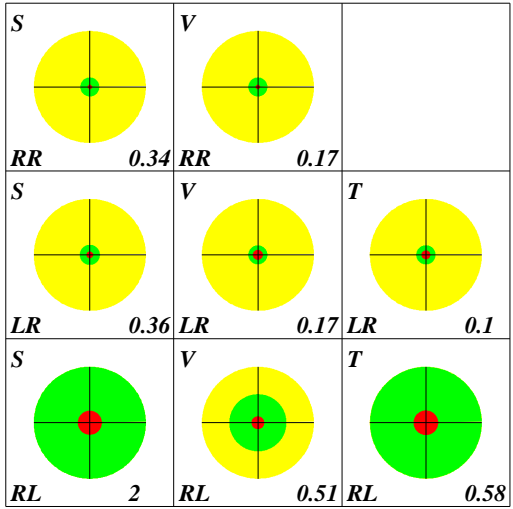}}
\caption{90\% CL limits \protect\cite{Boiko}
for the normalized $\tau$--decay couplings
$g'^n_{\epsilon\omega }\equiv g^n_{\epsilon\omega }/ N^n$,
where
$N^n \equiv \protect\mbox{\rm max}(|g^n_{\epsilon\omega }|) =2$,
1, $1/\protect\sqrt{3} $ for $n =$ S, V, T,
assuming $e/\mu$ universality.
The circles of unit area indicate the range allowed by the
normalization constraint (\protect\ref{eq:normalization}).
The experimental bounds are shown as shaded circles.
The $\mu $--decay limits are also shown (darker circles).}
\label{fig:tau_couplings}
\end{figure}
%%%%%%%%%%%%%%%%%%%%%%%%%%%%%%%%%%%

The experimental analysis of the $\tau$--decay parameters is
necessarily different from the one applied to the muon,
because of the much shorter $\tau$ lifetime.
The measurement of the $\tau$ polarization
is still possible due to the fact that the spins
of the $\tau^+\tau^-$ pair produced in $e^+e^-$ annihilation
are strongly correlated \cite{ABGPR:92,DDDR:93,FE:90}.  %BPR:91
Another possibility is to use the beam polarization, as done by SLD.
However, the polarization of the charged lepton emitted in the
$\tau$ decay has never been measured.
The experimental study of the inverse decay $\nu_\tau l^-\to\tau^-\nu_l$
looks far out of reach.

The determination of the $\tau$ polarization parameters
allows us to bound the total probability for the decay of
a right--handed $\tau$,
$Q_{\tau_R} \equiv Q_{RR} + Q_{LR}$.
At 90\% CL, one finds:
% (ignoring possible correlations among the measurements):
$Q_{\tau_R}^{\tau\to\mu} <  0.047$,
$Q_{\tau_R}^{\tau\to e} < 0.054$ and
$Q_{\tau_R}^{\tau\to l} < 0.032$,
where the last value refers to the $\tau$ decay into either
$l=e$ or $\mu$, assuming identical $e$/$\mu$ couplings.
These positive semidefinite probabilities imply corresponding limits
on all $|g^n_{RR}|$ and $|g^n_{LR}|$ couplings.
Including also the information from the energy distribution,
one gets the bounds shown in Fig.~\ref{fig:tau_couplings},
where $e$/$\mu$ universality has been assumed.

\section{SEARCHING FOR NEW PHYSICS}
\label{sec:new-physics}

\subsection{The Tau Neutrino}

The DONUT experiment at Fermilab has provided \cite{DONUT}
the first direct observation of the $\nu_\tau$
(produced through $p+N\to D_s + \cdots $,
followed by the decays $D_s\to\tau^-\bar\nu_\tau$
and $\tau^-\to\nu_\tau + \cdots $), through the detection
of $\nu_\tau + N\to\tau +X$. With this important achievement,
all SM fermions have been finally detected and the three--family
structure is definitely established.

The feasibility to detect $\tau$ neutrinos is of great importance,
in view of the recent SuperKamiokande results \cite{SKamioka}
suggesting $\nu_\mu\to\nu_\tau$ oscillations with
$m_{\nu_\tau}^2-m_{\nu_\mu}^2\sim (0.05\;\mbox{\rm eV})^2$.
This hypothesis could be corroborated making a long--baseline
neutrino experiment with a $\nu_\mu$ beam pointing into a far
($\sim 700$ Km) massive detector, able to detect the appearance of
a $\tau$ \cite{DiLella}.
%Several experimental proposals are being studied at present.

All observed $\tau$ decays are supposed to be accompanied by neutrino
emission, in order to fulfil energy--momentum conservation requirements.
From a two--dimensional likelihood fit of the
visible energy and the invariant--mass distribution of the final hadrons
in $\tau^-\to\nu_\tau X^-$ events, it is possible to set a limit on
the $\nu_\tau$ mass \cite{Duboscq}.
The strongest bound up to date \cite{ALEPHnumass},
\be\label{eq:numasslimit}
m_{\nu_\tau} \, < \, 18.2 \,\mbox{\rm MeV} \qquad (95\%\,
\mbox{\rm CL}),
\ee
has been obtained from a combined analysis of
$\tau^-\to (3 \pi)^-\nu_\tau, (5 \pi)^-\nu_\tau$
events.

\subsection{Lepton--Number Violation}

In the minimal SM with massless neutrinos, there is a
separately conserved additive lepton number for each generation.
All present data are consistent with this conservation law.
However, there are no strong theoretical
reasons forbidding a mixing among the  different  leptons, in the same
way as happens in the quark sector.
Many models in fact predict lepton--flavour or even
lepton--number violation at some level \cite{Ilakovac}.
Experimental searches for these processes
can provide information on the scale at which the new physics begins to
play a  significant role.

The present upper limits on lepton--flavour and
lepton--number violating decays of the $\tau$ \cite{CLEOlim} are
in the range of $10^{-5}$ to $10^{-6}$, which is far
away from the impressive bounds \cite{PDG:00} obtained in $\mu$ decay
[Br$(\mu^-\to e^- \gamma)  < 1.2 \times 10^{-11}$,
Br$(\mu^-\to e^- e^+ e^-) < 1.0 \times 10^{-12}$,
Br$(\mu^-\to e^-\gamma\gamma) <  7.2 \times 10^{-11} \, $ (90\% CL)].
Nevertheless, the $\tau$--decay limits start to put interesting
constraints on possible new physics contributions.
With future $\tau$ samples of $10^7$ events
per year, an improvement of two orders of magnitude would be possible.

\subsection{Dipole Moments}

Owing to their chiral--changing structure, the electroweak dipole
moments may provide important insights on the mechanism responsible
for mass generation. In general, one expects that a fermion of mass
$m_f$ (generated by physics at some scale $M\gg m_f$) will have
induced  dipole moments proportional to some power of $m_f/M$.
Therefore, heavy fermions such as the $\tau$ should be a good
testing ground for this kind of effects.
Of special interest \cite{Inami}
are the electric and weak dipole moments,
$d^{\gamma,Z}_\tau$, which  violate $T$ and $P$ invariance;
they constitute a good probe of CP violation.

The present experimental constraints on the electroweak dipole
moments of the $\tau$ have been recently reanalyzed, using
effective operator techniques \cite{GSV:00}.
The achieved sensitivity is still marginal, but it is approaching
the level of the SM contribution to the $\tau$
anomalous magnetic moment \cite{NA:78}:
$a^\gamma_\tau\big |_{\mbox{\rms th}}  =  (1.1773 \pm 0.0003)
  \times 10^{-3}$.

\section{HADRONIC DECAYS}
\label{sec:hadronic}

%%%%%%%%%%%%%%%%%%%%%%%%%%%%%%%%%%%%%%%%%%%%%%
%%
%%   FIGURE Pion Form Factor
%%
\begin{figure}[tbh]
\myfigure{\epsfxsize =7.5cm \epsfbox{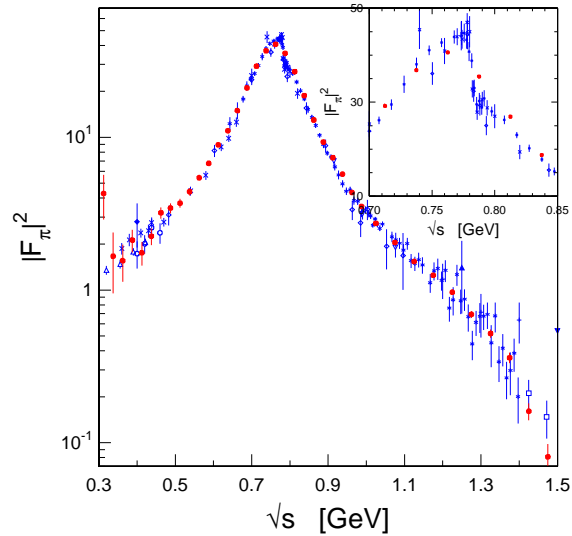}}
\vspace{-0.5cm}
\caption{Pion form factor from $\tau^-\to\nu_\tau\pi^-\pi^0$
\protect\cite{CLEOpiff} (filled circles) and
$e^+e^-\to\pi^+\pi^-$ data.}
\label{fig:pionff}
\end{figure}
%%%%%%%%%%%%%%%%%%%%%%%%%%%%%%%%%%%%%%%%%%%%%%%

The semileptonic decay modes $\tau^-\to\nu_\tau H^-$ probe the  matrix
element of the left--handed charged current between the vacuum and the
final hadronic state $H^-$.

For the decay modes with lowest multiplicity,
$\tau^-\to\nu_\tau\pi^-$  and $\tau^-\to\nu_\tau K^-$, the  relevant
matrix  elements  are  already  known  from  the  measured  decays
$\pi^-\to\mu^-\bar\nu_\mu$  and  $K^-\to\mu^-\bar\nu_\mu$.
The corresponding $\tau$ decay widths can then be predicted
rather accurately.
As shown in Table~\ref{tab:ccuniv}, these predictions are in good
agreement with the measured values, and provide a quite precise test
of charged--current universality.

For the two--pion final state, the hadronic matrix element is
parameterized in terms of the so-called pion form factor:
\bel{eq:Had_matrix}
\langle \pi^-\pi^0| \bar d \gamma^\mu  u | 0 \rangle \equiv
\sqrt{2}\, F_\pi(s)\, \left( p_{\pi^-}- p_{\pi^0}\right)^\mu \, .
\ee
Figure \ref{fig:pionff} shows the recent CLEO measurement
of $|F_\pi(s)|^2$ from $\tau\to\nu_\tau\pi^-\pi^0$ data \cite{CLEOpiff}
(a similar analysis was done previously by ALEPH \cite{ALEPHpiff}).
Also shown is the corresponding determination from
$e^+e^-\to\pi^+\pi^-$ data.
The precision achieved with $\tau$ decays is clearly better.
There is good agreement between both sets of data, although
the $\tau$ points tend to be slightly higher \cite{Eidelman}.

A dynamical understanding of the pion form factor can be achieved
\cite{PP:97,DPP:00,Portoles},
by using analyticity, unitarity and some general properties of QCD.

At low momentum transfer, the coupling of any
number of $\pi $'s, $K$'s and $\eta$'s to the
V$-$A current can be rigorously calculated with
Chiral Perturbation Theory \cite{GL:85,EC:95} techniques,
as an expansion in powers of $s$ and light quark masses over the
chiral symmetry breaking scale ($\Lambda_\chi\sim 1$ GeV).
This includes chiral loop corrections, which encode the absorptive
contributions required by unitarity.
The short--distance information is contained in the so-called
chiral couplings, which are known to be dominated by the effect
of the lowest--mass resonances \cite{EGPR:89}.

%%%%%%%%%%%%%%%%%%%%%%%%%%%%%%%%%%%%%%%%%%%%%%
%%
%%   FIGURE Pion Form Factor
%%
\begin{figure}[tbh]
\myfigure{\rotate[r]{\epsfysize =7.2cm \epsfbox{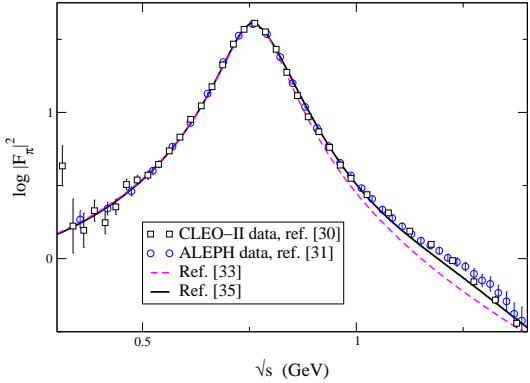}}}
\vspace{-0.5cm}
\caption{Pion form factor data compared with theoretical
predictions \protect\cite{Portoles}.}
\label{fig:pionth}
\end{figure}
%%%%%%%%%%%%%%%%%%%%%%%%%%%%%%%%%%%%%%%%%%%%%%%

In the limit of an infinite number of quark colours $N_C$, QCD reduces
to a theory of tree--level resonance exchanges \cite{HO:74}.
Thus, the $\rho$ propagator governs the pion form factor
at $\sqrt{s}\lsim 1$ GeV, providing an all-order
resummation of the polynomic chiral corrections.
Moreover, requiring
$F_\pi(s)$ to satisfy the correct QCD behaviour at large $s$,
one can determine the relevant $\rho$ couplings \cite{EGPR:89}.
The leading $1/N_C$ corrections correspond to
pion loops and can be incorporated by matching the
large--$N_C$ result with the Chiral Perturbation Theory description
\cite{PP:97}. Using analyticity and unitarity constraints,
the chiral logarithms associated with those pion loops
can be exponentiated to all orders in the chiral expansion. 
Putting all these
fundamental ingredients together, one gets the result \cite{PP:97}:
$$
F_\pi(s) = {M_\rho^2\over M_\rho^2 - s - i M_\rho \Gamma_\rho(s)}
\exp{\left\{-{s \,\mbox{\rm Re} A(s)           %\left[A(s)\right]
\over 96\pi^2f_\pi^2} \right\}} ,
$$
where
$$
A(s) \equiv \log{\left({m_\pi^2\over M_\rho^2}\right)} +
8 {m_\pi^2 \over s} - {5\over 3} +
\sigma_\pi^3 \log{\left({\sigma_\pi+1\over \sigma_\pi-1}\right)}
$$
contains the one-loop chiral logarithms,
$\sigma_\pi\equiv\sqrt{1-4m_\pi^2/s}$ and the
off-shell $\rho$ width is given by \cite{PP:97,DPP:00}
$\Gamma_\rho(s) = \theta(s-4m_\pi^2)\,\sigma_\pi^3\,
M_\rho s/(96\pi f_\pi^2)$.
This prediction, which only depends on $M_\rho$,
$m_\pi$ and the pion decay constant
$f_\pi$, is compared with the data
in Fig.~\ref{fig:pionth}. The agreement is
rather impressive and extends to negative $s$ values, where
the $e^-\pi$ elastic data (not shown in the figure) sits.

One can include the effect of higher $\rho$ resonances, at the price
of having some free parameters (subtraction constants)
which decrease the predictive power \cite{Portoles}.
This gives a better description of the $\rho'$ shoulder around 1.2 GeV.

The dynamical structure of other hadronic final states has been also
investigated. CLEO has measured \cite{CLEO3pi}
the four $J^P=1^+$ structure functions
characterizing the decay $\tau^-\to\nu_\tau\pi^- 2\pi^0$,
improving a previous OPAL analysis \cite{OPAL3pi}.
A theoretical analysis of these data is in progress \cite{DPP:01}.

\section{THE TAU HADRONIC WIDTH}
\label{sec:hadronic_width}

The inclusive character of the total $\tau$ hadronic width
renders possible an accurate calculation of the ratio
\cite{BR:88,NP:88,BNP:92,LDP:92a,QCD:94}
%[$(\gamma)$ represents additional photons or lepton pairs]
%
\bel{eq:r_tau_def}
R_\tau \equiv { \Gamma [\tau^- \rightarrow \nu_\tau
                   \,\mbox{\rm hadrons}\, (\gamma)] \over
                         \Gamma [\tau^- \rightarrow
                \nu_\tau e^- {\bar \nu}_e (\gamma)] } ,
\ee
using analyticity constraints and the Operator Product Expansion
(OPE).

The theoretical analysis of $R_\tau$ involves
the two--point correlation functions
\bel{eq:pi_j}
\Pi^{\mu\nu}_j(q)\equiv i \int d^4x \; e^{iqx}\;
\langle 0|T(j^\mu(x) j^\nu(0)^\dagger)|0\rangle
\ee
for the vector,
$j^\mu = V^{\mu}_{ij} \equiv \bar{\psi}_j \gamma^{\mu} \psi_i$,
and axial--vector,
$j^\mu = A^{\mu}_{ij} \equiv \bar{\psi}_j \gamma^{\mu} \gamma_5 \psi_i$,
colour--singlet quark currents ($i,j=u,d,s$).
They have the Lorentz decompositions
\beqn\label{eq:lorentz}
\Pi^{\mu \nu}_{ij,V/A}(q) & \!\!\!\! = & \!\!\!\!
  (-g^{\mu\nu} q^2 + q^{\mu} q^{\nu}) \, \Pi_{ij,V/A}^{(1)}(q^2)   \no\\
  && \!\!\!\! +   q^{\mu} q^{\nu} \, \Pi_{ij,V/A}^{(0)}(q^2) ,
\eeqn
where the superscript $J=0,1$ %$J=1$ or $J=0$
denotes the angular momentum in the hadronic rest frame.

The imaginary parts of the two--point functions
$\, \Pi^{(J)}_{ij,V/A}(q^2) \, $
are proportional to the spectral functions for hadrons with the
corresponding quantum numbers.  The hadronic decay rate of the $\tau$
can be written as an integral of these spectral functions
over the invariant mass $s$ of the final--state hadrons:
\beqn\label{eq:spectral}
R_\tau  &\!\!\!\!\! = &\!\!\!\!\!
12 \pi \int^{m_\tau^2}_0 {ds \over m_\tau^2 } \,
 \left(1-{s \over m_\tau^2}\right)^2
\\ &\!\!\!\!\! \times &\!\!\!\!\!
\biggl[ \left(1 + 2 {s \over m_\tau^2}\right)
 \mbox{\rm Im} \Pi^{(1)}(s)
 + \mbox{\rm Im} \Pi^{(0)}(s) \biggr]  .\no
\eeqn
 The appropriate combinations of correlators are
\beqn\label{eq:pi}
\Pi^{(J)}(s)  &\!\!\! \equiv  &\!\!\!
  |V_{ud}|^2 \, \left( \Pi^{(J)}_{ud,V}(s) + \Pi^{(J)}_{ud,A}(s) \right)
\no\\ &\!\!\! + &\!\!\!
|V_{us}|^2 \, \left( \Pi^{(J)}_{us,V}(s) + \Pi^{(J)}_{us,A}(s) \right).
\eeqn

We can separate the inclusive contributions associated with
specific quark currents:
\be\label{eq:r_tau_v,a,s}
 R_\tau \, = \, R_{\tau,V} + R_{\tau,A} + R_{\tau,S}\, .
\ee
$R_{\tau,V}$ and $R_{\tau,A}$ correspond to the first two terms
in \eqn{eq:pi}, while
$R_{\tau,S}$ contains the remaining Cabibbo--suppressed contributions.
Non-strange hadronic decays of the $\tau$ are resolved experimentally
into vector ($R_{\tau,V}$) and axial-vector ($R_{\tau,A}$)
contributions according to whether the
hadronic final state includes an even or odd number of pions.
Strange decays ($R_{\tau,S}$) are of course identified by the
presence of an odd number of kaons in the final state.

Since the hadronic spectral functions are sensitive to the non-perturbative
effects of QCD that bind quarks into hadrons, the integrand in
Eq.~\eqn{eq:spectral} cannot be calculated at present from QCD.
Nevertheless the integral itself can be calculated systematically
by exploiting
the analytic properties of the correlators $\Pi^{(J)}(s)$.
They are analytic
functions of $s$ except along the positive real $s$--axis, where their
imaginary parts have discontinuities.
%The integral \eqn{eq:spectral}
$R_\tau$ can
therefore be expressed as a contour integral
in the complex $s$--plane running
counter-clockwise around the circle $|s|=m_\tau^2$:
\beqn\label{eq:circle}
 R_\tau &\!\!\!\!\! =&\!\!\!\!\!
6 \pi i \oint_{|s|=m_\tau^2} {ds \over m_\tau^2} \,
 \left(1 - {s \over m_\tau^2}\right)^2
\\ &\!\!\!\!\!\times &\!\!\!\!\!
%\qquad\no \\ &\!\!\! &\!\!\!\!\!\!\!\!\!\!\!\! \!\!\!\times
 \left[ \left(1 + 2 {s \over m_\tau^2}\right) \Pi^{(0+1)}(s)
         - 2 {s \over m_\tau^2} \Pi^{(0)}(s) \right]  . \no
\eeqn
This expression requires the correlators only for
complex $s$ of order $m_\tau^2$, which is significantly larger than
the scale
associated with non-perturbative effects in QCD.  The short--distance
OPE can therefore be used to organize
the perturbative and non-perturbative contributions
to the correlators into a systematic expansion \cite{SVZ:79}
in powers of $1/s$.
 The possible uncertainties associated with the use of the OPE near the
 time-like axis are negligible in this case, because
 the integrand in \eqn{eq:circle} includes a factor
 $(1- s/m_\tau^2)^2$, which provides a double zero at $s=m_\tau^2$,
 effectively suppressing the contribution from the
 region near the branch cut.

%%%%%%%%%%%%%%%%%% FIGURE %%%%%%%%%%%%%%%%%%%%%
\begin{figure}[tb]
\label{fig:circle}
\myfigure{\epsfxsize =6.5cm \epsfbox{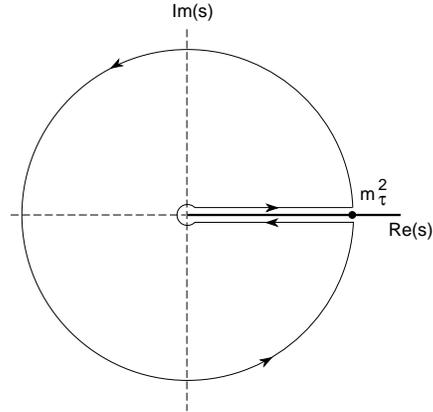}}
\vspace{-0.5cm}
\caption{Integration contour in the complex s--plane, used to obtain
Eq.~\protect\eqn{eq:circle}}
\end{figure}
%%%%%%%%%%%%%% END FIGURE %%%%%%%%%%%%%%%%%%%%%

After evaluating the contour integral, $R_\tau$
can be expressed as an expansion in powers of $1/m_\tau^2$,
with coefficients that depend only logarithmically on $m_\tau$:
\bel{eq:r_total}
R_{\tau} = 3\,
S_{EW} \left\{ 1 + \delta_{EW}'  %+ \delta^{(0)}
+ \sum_{D=0,2,...} \delta^{(D)}\right\} .
\ee
The factors $S_{EW}=1.0194$ and $\delta_{EW}'=0.0010$
contain the known electroweak corrections at the leading
\cite{MS:88} and next-to-leading \cite{BL:90} logarithm
approximation.
The dimension--0 contribution,
$\delta^{(0)}$,
is the purely perturbative correction neglecting quark masses.
%which is the same for all the components of $R_\tau$.
It is given by
\cite{BR:88,NP:88,BNP:92,LDP:92a,QCD:94}:
\beqn\label{eq:delta0}
\delta^{(0)}  &\!\!\! =&\!\!\! \sum_{n=1}  K_n \, A^{(n)}(\alpha_s)
\no\\ &\!\!\! = &\!\!\!
a_\tau + 5.2023 \, a_\tau^2 + 26.366 \, a_\tau^3
       + \, \cO(a_\tau^4)  \, ,
\eeqn
where $a_\tau\equiv\alpha_s(m_\tau^2)/\pi$.

The dynamical coefficients $K_n$ regulate the perturbative expansion
of  $ -s{d\over ds}\Pi^{(0+1)}(s)$
in the massless--quark limit
[$s\,\Pi^{(0)}(s)=0$ for massless quarks]; they are
known \cite{ChKT:79,GKL:91} to $\cO(\alpha_s^3)$:
$K_1 = 1$; $K_2 = 1.6398$; $K_3(\overline{MS}) = 6.3710$.
The kinematical effect of the contour integration is contained in
the functions \cite{LDP:92a}
\beqn\label{eq:a_xi}
A^{(n)}(\alpha_s) &\!\!\!\! = &\!\!\!\! {1\over 2 \pi i}
\oint_{|s| = m_\tau^2} {ds \over s} \,
  \left({\alpha_s(-s)\over\pi}\right)^n
\no\\ &\!\!\!\!\times  &\!\!\!\!
 %\!\!\!\!\!\!\!\!\!\!\!\! \times
 \left( 1 - 2 {s \over m_\tau^2} + 2 {s^3 \over m_\tau^6}
         - {s^4 \over  m_\tau^8} \right) ,
\eeqn
which only depend on $\alpha_s(m_\tau^2)$.
Owing to the long running of the strong coupling along the circle, the
coefficients of the
perturbative expansion of $\delta^{(0)}$ in powers of
$\alpha_s(m_\tau^2)$ are larger than the direct $K_n$
contributions. This running effect can be properly resummed to all orders
in $\alpha_s$ by fully keeping \cite{LDP:92a}
the known four--loop--level calculation of
the integrals $A^{(n)}(\alpha_s)$.

The leading quark--mass corrections $\delta^{(2)}$
are tiny for the up and down quarks.
The correction from the strange quark mass is important
for strange decays but, owing
to the $|V_{us}|^2$ suppression, the
effect on the total ratio $R_{\tau}$ is below 1\%.

The non-perturbative contributions can be shown to be
suppressed by six powers of the $\tau$ mass \cite{BNP:92}
and, therefore, are very small. Their numerical size has been
determined from the invariant--mass distribution of the final hadrons
in $\tau$ decay, through the study of weighted integrals \cite{LDP:92b},
\be
R_{\tau}^{kl} \equiv \int_0^{m_\tau^2} ds\,
\left(1 - {s\over m_\tau^2}\right)^k\, \left({s\over m_\tau^2}\right)^l\,
{d R_{\tau}\over ds} \, ,
\ee
which can be calculated theoretically in the same way as $R_{\tau}$.
The predicted suppression \cite{BNP:92}
of the non-perturbative corrections has been confirmed by
ALEPH \cite{ALEPH:98}, CLEO \cite{CLEO:95} and OPAL \cite{OPAL:98}.
The most recent analyses \cite{ALEPH:98,OPAL:98} give
\bel{eq:del_np}
\delta_{\mbox{\rms NP}} \equiv \sum_{D\ge 4} \delta^{(D)} =
-0.003\pm 0.003 \, .
\ee
%

%%%%%%%%%%%%%%%%%% FIGURE %%%%%%%%%%%%%%%%%%%%%
%              alpha_s running
%
\begin{figure}[tbh]
\label{fig:alpha_s}
\myfigure{\epsfxsize =7.5cm \epsfbox{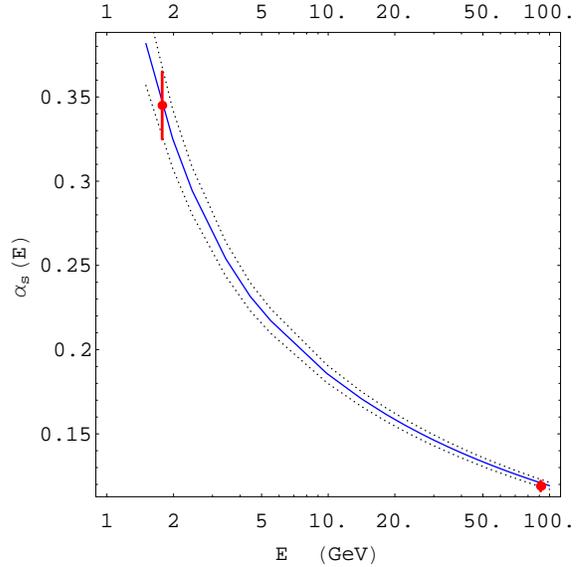}}
\vspace{-0.5cm}
\caption{Measured values of $\alpha_s$ in $\tau$ and $Z$ decays.
The curves show the energy dependence predicted by QCD, using
$\alpha_s(m_\tau^2)$ as input.}
\end{figure}
%%%%%%%%%%%%%% END FIGURE %%%%%%%%%%%%%%%%%%%%%

The QCD prediction
for $R_{\tau,V+A}$ is then completely dominated by the
perturbative contribution $\delta^{(0)}$;
non-perturbative effects being smaller
than the perturbative uncertainties from uncalculated higher--order
corrections. The result turns out to be
very sensitive to the value of $\alpha_s(m_\tau^2)$, allowing for an accurate
determination of the fundamental QCD coupling.
The experimental measurement \cite{ALEPH:98,OPAL:98}
$R_{\tau,V+A}= 3.484\pm0.024$ implies $\delta^{(0)} = 0.200\pm 0.013$,
which corresponds (in the $\overline{\rm MS}$ scheme) to
\be\label{eq:alpha}
\alpha_s(m_\tau^2)  =  0.345\pm 0.020 \, .
\ee

The strong coupling measured at the $\tau$ mass
scale is significatively different from the values obtained at higher energies.
From the hadronic decays of the $Z$ boson, one gets
$\alpha_s(M_Z^2) = 0.119\pm 0.003$ \cite{PDG:00,LEP:99}, 
which differs from the $\tau$ decay
measurement by eleven standard deviations.
After evolution up to the scale $M_Z$ \cite{Rodrigo:1998zd},
the strong coupling constant in \eqn{eq:alpha} decreases to
\be\label{eq:alpha_z}
\alpha_s(M_Z^2)  =  0.1208\pm 0.0025 \, ,
\ee
in excellent agreement with the direct measurements at the $Z$ peak
and with a similar accuracy.
The comparison of these two determinations of $\alpha_s$ in two extreme
energy regimes, $m_\tau$ and $M_Z$, provides a beautiful test of the
predicted running of the QCD coupling;
i.e. a very significant experimental verification of {\it asymptotic freedom}.

%%%%%%%%%%%%%%%%%% FIGURE %%%%%%%%%%%%%%%%%%%%%
%             Spectral Functions
%
\begin{figure}[tbh]
\myfigure{\epsfysize =7cm \epsfbox{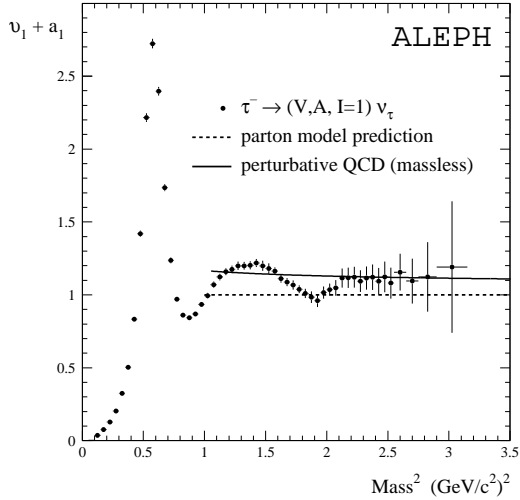}}
\vspace{-0.5cm}
\caption{$V+A$ spectral function \protect\cite{ALEPH:98}.}
\label{fig:V+Asf}
\end{figure}
\begin{figure}[tbh]
\myfigure{\epsfysize =7cm \epsfbox{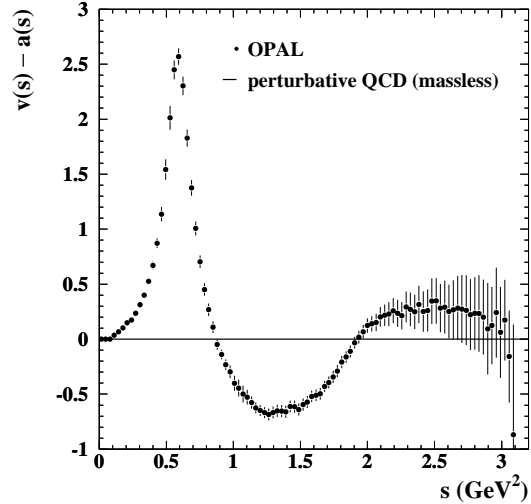}}
\vspace{-0.5cm}
\caption{$V-A$ spectral function \protect\cite{OPAL:98}.}
\label{fig:V-Asf}
\end{figure}
%%%%%%%%%%%%%% END FIGURE %%%%%%%%%%%%%%%%%%%%%

From a careful analysis of the hadronic invariant--mass distribution,
ALEPH \cite{ALEPHpiff,ALEPH:98} and OPAL \cite{OPAL:98} have measured
the spectral functions associated with the vector and axial--vector
quark currents. Their difference is a pure non-perturbative quantity,
which carries important information on the QCD dynamics
\cite{BNP:92,WE:67,PhV-A,KPR:98,ThV-A};
it allows to determine low--energy parameters, such
as the pion decay constant, the electromagnetic pion mass difference
$m_{\pi^\pm}-m_{\pi^0}$, or the axial pion form factor, 
in good agreement with their direct measurements \cite{Davier}.

The vector spectral function has been also used to measure the hadronic
vacuum polarization
effects associated with the photon and, therefore, estimate
how the electromagnetic fine structure constant
gets modified at LEP energies.
The uncertainty of this parameter is one
of the main limitations in the extraction of the Higgs mass from global
electroweak fits to the LEP/SLD data.
From the ALEPH $\tau$ data \cite{ALEPHpiff},
the Orsay group obtains \cite{Davier,orsay}
$\alpha^{-1}(M_Z) = 128.933 \pm 0.021$, which reduces the error
of the fitted $\log{(M_H)}$ value by 30\%.
The same $\tau$ data allows to pin down the hadronic contribution to
the anomalous magnetic moment of the muon \index{anomalous magnetic moment}
$a^\gamma_\mu$. The recent       %ALEPH \cite{orsay} and CLEO
analyses \cite{CLEOpiff,orsay} have improved the theoretical
prediction of $a^\gamma_\mu$,
setting a reference value  to be compared with the
measurement of the
BNL-E821 experiment, presently running at Brookhaven.

\section{THE STRANGE QUARK MASS}

%%%%%%%%%%%%%%%%%% FIGURE %%%%%%%%%%%%%%%%%%%%%
%            S=1 Spectral Function
%
\begin{figure}[tbh]
\label{fig:S=1sf}
\myfigure{\epsfxsize =7.5cm \epsfbox{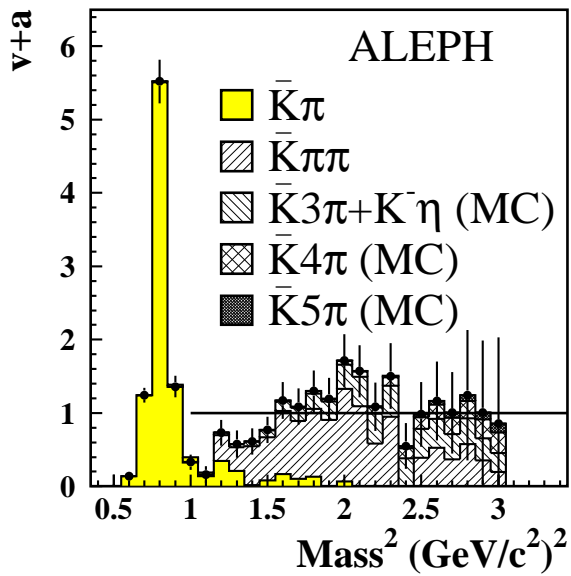}}
\vspace{-0.5cm}
\caption{$|\Delta S|=1$ spectral function \protect\cite{ALEPHms}.}
\end{figure}
%%%%%%%%%%%%%% END FIGURE %%%%%%%%%%%%%%%%%%%%%

The LEP experiments and CLEO have performed an extensive investigation
of kaon production in $\tau$ decays \cite{kaons,ALEPHms}.
ALEPH has determined the inclusive invariant mass
distribution of the final hadrons in the Cabibbo--suppressed decays
\cite{ALEPHms}.
The separate measurement of the $|\Delta S|=0$ and $|\Delta S|=1$
decay widths allows us to pin down the SU(3) breaking effect induced
by the strange quark mass, through the differences
\beqn
\lefteqn{\delta R_\tau^{kl}  \equiv
  {R_{\tau,V+A}^{kl}\over |V_{ud}|^2} - {R_{\tau,S}^{kl}\over |V_{us}|^2}
  } &&
\\ & &\approx  24\, {m_s^2(m_\tau^2)\over m_\tau^2} \, \Delta_{kl}(\alpha_s)
    - 48\pi^2\, {\delta O_4\over m_\tau^4} \, Q_{kl}(\alpha_s)
\, ,\no
\eeqn
where $\Delta_{kl}(\alpha_s)$ and $Q_{kl}(\alpha_s)$ are perturbative QCD
corrections, which are known to $O(\alpha_s^3)$ and $O(\alpha_s^2)$,
respectively \cite{PP:99}.
The small non-perturbative contribution,
$\delta O_4 \equiv\langle 0| m_s \bar s s - m_d \bar d d |0\rangle
 = -(1.5\pm 0.4)\times 10^{-3}\;\mbox{\rm GeV}^4$,
has been estimated with Chiral Perturbation Theory techniques \cite{PP:99}.
Table~\ref{tab:ms}
shows the measured \cite{ALEPHms,CH:00} differences
$\delta R_\tau^{kl}$ and the corresponding ($\overline{\rm MS}$)
values \cite{CH:00} of $m_s(m_\tau^2)$.
The theoretical errors are dominated by the very
large perturbative uncertainties of $\Delta_{kl}(\alpha_s)$
\cite{PP:99,CK:93,CH:97,MA:98,ChKP:98}.

%%%%%%%%%%%%%%%%%%% TABLE M_S %%%%%%%%%%%%%%%%%%
%%
\begin{table}[tb]
\caption{Measured moments $\delta R_\tau^{kl}$
\protect\cite{ALEPHms,CH:00}  and corresponding $m_s(m_\tau^2)$ values
\protect\cite{CH:00}.}
\label{tab:ms}
\begin{tabular}{c|c|c}
$(k,l)$ & $\delta R_\tau^{kl}$ & $m_s(m_\tau^2)$ \ (MeV)
\\ \hline
$(0,0)$ & $0.370\pm 0.133$ & $131\pm 29_{\rms exp}\pm 14_{\rms th}$
\\
$(1,0)$ & $0.396\pm 0.078$ & $119\pm 16_{\rms exp}\pm 12_{\rms th}$
\\
$(2,0)$ & $0.397\pm 0.054$ & $104\pm 11_{\rms exp}\pm 19_{\rms th}$
\end{tabular}
\end{table}
%%%%%%%%%%%%%%%%%%%%%%%%%%%%%%%%%%%%%%%%%%%%%%%%%

A global analysis, using the information from the three moments and
taking into account the strong error correlations,
gives the result \cite{CH:00}
$$
m_s(m_\tau^2) = \left(112 \pm 23\right)\;\mbox{\rm MeV} \, .
$$
This corresponds to
$m_s(1\:\mbox{\rm GeV}^2) = (150\pm 35)$ MeV.
A similar result is obtained from an analysis based on ``optimal moments'',
with improved perturbative convergence \cite{Maltman}.

Previous estimates of $m_s$ were based on  lattice simulations or
phenomenological QCD sum rules. There is a rather large spread of
lattice results \cite{lattice}; the average value agrees with the $\tau$
determination, but some results are too small and could be in conflict with QCD 
lower bounds \cite{Bounds}. 
The latest QCD sum rules \cite{QCDSR} results are
compatible with the $\tau$ value. The advantage of the $\tau$ determination
is the direct use of experimental input, which makes easier to quantify
the associated uncertainties.

\section{SUMMARY}
\label{sec:summary}

The flavour structure of the SM is one of the main pending questions
in our understanding of weak interactions. Although we do not know the
reason of the observed family replication, we have learned experimentally
that the number of SM fermion generations is just three (and no more).
Therefore, we must study as precisely as possible the few existing flavours
to get some hints on the dynamics responsible for their observed structure.

The $\tau$ turns out to be an ideal laboratory to test the SM.
It is a lepton, which means clean physics, and moreover it is
heavy enough to produce a large variety of decay modes.
Na\"{\i}vely, one would expect the $\tau$ to be much more sensitive
than the $e$ or the $\mu$ to new physics related to the flavour and
mass--generation problems.

QCD studies can also benefit a lot from the existence of this heavy lepton,
able to decay into hadrons. Owing to their semileptonic character, the
hadronic $\tau$ decays provide a powerful tool to investigate the low--energy
effects of the strong interactions in rather simple conditions.

Our knowledge of the $\tau$ properties has been considerably
improved during the last few years.
Lepton universality has been tested to rather good accuracy,
both in the charged and neutral current sectors. The
Lorentz structure of the leptonic $\tau$ decays is certainly not determined,
but begins to be experimentally explored. 
An upper limit of 3.2\% (90\% CL) has been already set on the probability
of having a (wrong) decay from a right--handed $\tau$.
The quality of the hadronic data
has made possible to perform quantitative QCD tests
and determine the strong coupling constant very accurately.
Searches for non-standard phenomena have been pushed to the limits
that the existing data samples allow to investigate.

At present, all experimental results on the $\tau$ lepton are consistent with
the SM. There is, however, large room for improvements. Future $\tau$ experiments
will probe the SM to a much deeper level of sensitivity and will explore the
frontier of its possible extensions.

\section*{ACKNOWLEDGEMENTS}
I thank the organizers for hosting an enjoyable conference and
J. Portol\'es for useful discussions.
This work has been supported in part by the ECC, TMR Network EURODAPHNE
(ERBFMX-CT98-0169), and by DGESIC (Spain) under grant No. PB97-1261.

%%%%%%%%%%%%%%%%%%%%%%%%%%%%%%%%%%%%%%

\end{document}